\documentstyle[11pt]{article}
\newcommand{\SC}{\langle{\cal O}_8^{\psi '}(^1S_0)\rangle}
\newcommand{\SD}{\langle{\cal O}_8^{\psi '}(^3S_1)\rangle}
\newcommand{\PS}{\langle{\cal O}_8^{\psi '}(^3P_0)\rangle}

\begin{document}

\baselineskip=21pt

\begin{flushright}
KOBE--FHD--97--01\\
FUT--TH--97--01~~~\\
May~~~~~~~~1997
\end{flushright}
\begin{center}
{\Large\bf Hard Processes with Spin--Dependent Parton Distributions}\\
\vspace{10mm}
T.\ Yamanishi\footnote[2]{E--mail: yamanisi@natura.h.kobe-u.ac.jp~,~~
Phone: +81--6--879--8940~,~~Fax: +81--6--879--8899~.},
T.\ Morii$^1$, D.\ Roy$^1$, K.\ Sudoh$^2$ and S.\ Tanaka$^3$\\
\vspace{5mm}
{\it Research Center for Nuclear Physics,
Osaka University, Ibaraki, Osaka, 567, Japan}\\
{\it $^1$Faculty of Human Development, Kobe University, Nada, Kobe 657,
Japan}\\
{\it $^2$Department of Education, Kobe University, Nada, Kobe 657,
Japan}\\
{\it $^3$Department of Physics, Hyogo University of Education,
Yashiro--cyo, Hyogo 673--14, Japan}
\end{center}

\begin{center}
ABSTRACT

\vspace{5mm}
\begin{minipage}{130 mm}
To test a possible color--octet contribution of heavy quarkonium productions,
we propose $\psi$'--productions for small--$p_T$ regions, where
the color--singlet $\psi$' state cannot be produced by the gluon--gluon
fusion owing to charge conjugation.
We also point out that these processes are very useful to deduce
informations of spin--dependent gluon distributions $\Delta g$. 
\end{minipage}
\end{center}

\vspace{5mm}

Recently, it has been reported that the cross sections of heavy quarkonium
productions measured in unpolarized $p\bar p$ collisions by
the CDF collaboration are inconsistent with the calculation with the lowest
order processes alone\cite{CDF}.
Several groups have proposed new mechanisms: the production proceeds
with not only color--singlet but also color--octet quarkonim states and
this new mechanism dominates over the lowest order processes for
large--$p_T$ quarkonium productions\cite{octets}.
But it does not seem to work for $\gamma+p\to J/\psi+X$\cite{Cacciari96}.
The discussion on the color--octet contribution is still controversial.
Thus, it is important to test this mechanism in other experiments.

On the other hand, measurements of the spin--dependent structure functions
of a nucleon $g_1(x, Q^2)$ for polarized deep inelastic scatterings (PDIS)
imply a surprizing result that the amount of the proton spin carried
by quarks, ($\sim 30$\%), is much smaller than expected from the
quark--parton model\cite{pDIS}.
One of keys to solve this spin puzzle is to directly measure the
spin--dependent gluon distributions $\Delta g$ and study the behavior of
them.
Relativistic Heavy Ion Collider (RHIC) at Brookhaven National Laboratory
(BNL), which will start early in the next decade, could offer such a
possibility and will give us fruitful informations on the nucleon spin
structure.

In this talk, to test a possible color--octet contribution to charmonium
productions\cite{octets} and furthermore to deduce informations of the
spin--dependent gluon distributions, we suggest $\psi$'--productions for
small--$p_T$ regions at RHIC energies, where the color--singlet $\psi$'
state does not contribute to the process due to charge conjugation.
This process proceeds via the gluon--gluon fusion in the lowest order.
Although there are several charmonium states, the detection for
$\psi$' is easier than that for other charmonium states.

Let us introduce a two--spin asymmetry $A_{LL}$ for this process,
\begin{equation}
A_{LL} = \frac{\left[d\sigma_{++}-d\sigma_{+-}+
d\sigma_{--}-d\sigma_{-+}\right]}
{\left[d\sigma_{++}+d\sigma_{+-}+
d\sigma_{--}+d\sigma_{-+}\right]} = \frac{d\Delta\sigma}{d\sigma}~,
\label{eqn:A_LL}
\end{equation}
where $d\sigma_{+-}$, for instance, denotes that the helicity
of one beam particle is positive and the other is negative.

The spin--dependent and spin--independent differential cross sections are
\begin{eqnarray}
\frac{d\Delta\sigma}{dx_L}&=&
\frac{\tau}{\sqrt{x_L^2+4\tau}}\left[\frac{\pi^3\alpha_s^2}{144m_c^5}
\{\SC-\frac{1}{m_c^2}\PS\}
\Delta g(x_a, Q^2)\Delta g(x_b, Q^2)\right .\nonumber\\
&&\left .-\frac{\pi^3\alpha_s^2}{54m_c^5}\SD
\{\Delta q(x_a, Q^2)\Delta\bar q(x_b, Q^2)+
\Delta q\leftrightarrow\Delta\bar q\} \right]~,
\label{eqn:dDs2}\\
\frac{d\sigma}{dx_L}&=&
\frac{\tau}{\sqrt{x_L^2+4\tau}}\left[\frac{\pi^3\alpha_s^2}{144m_c^5}
\{\SC+\frac{7}{m_c^2}\PS\} g(x_a, Q^2)~g(x_b, Q^2)\right .\nonumber\\
&&\left .+\frac{\pi^3\alpha_s^2}{54m_c^5}\SD
\{q(x_a, Q^2)~\bar q(x_b, Q^2)+q\leftrightarrow\bar q\}\right]~,
\label{eqn:ds2}
\end{eqnarray}
where $x_a$ and $x_b$ are the momentum fraction in a proton and given as
\begin{equation}
x_a=\frac{x_L+\sqrt{x_L^2+4\tau}}{2}~,~~
x_b=\frac{-x_L+\sqrt{x_L^2+4\tau}}{2}~,~~
x_L\equiv \frac{2p_L}{\sqrt s}~,~~
\tau\equiv \frac{M^2}{s}~,
\label{eqn:def-x}
\end{equation}
with longitudinal momentum $p_L$ of the produced particle.
$\SC$, $\SD$ and $\PS$ are nonperturbative long--distance factors
associated with the production of a $c\bar c$ pair in a color--octet
$^1S_0$, $^3S_1$ and $^3P_0$ states, respectively. From
recent analysis for charmonium hadroproductions, the value of $\SD$
and of the combination are given as
$\SD\sim 4.6\times 10^{-3}~{\rm [GeV^3]}$,
$\SC+\frac{7}{m_c^2}\PS\sim 5.2\times 10^{-3}~{\rm [GeV^3]}$ \cite{Beneke96},
and also another combination
$\frac{1}{3}\SC+\frac{1}{m_c^2}\PS\sim (5.9\pm 1.9)\times
10^{-3}~{\rm [GeV^3]}$ \cite{Leibovich96}.
Then we find the ratio as
$\frac{\tilde\Theta}{\Theta}\equiv
\frac{\SC-\frac{1}{m_c^2}\PS}{\SC+\frac{7}{m_c^2}\PS}
\sim 0.69\div 1.63$.

In order to study how the two--spin asymmetry $A_{LL}^{\psi '}$ is affect
by the spin--dependent gluon distribution $\Delta g(x)$, we take a model of
$\Delta g(x)$.
So far, many people have suggested various kinds of $\Delta g(x)$ from
the analysis of the data on $g_1(x,Q^2)$\cite{GS95,BBS95,GRV95}.
The $x$--dependence of $x\Delta g(x, Q^2)$ at $Q^2=10$GeV$^2$ is shown
in fig.1.
We estimate the $A_{LL}^{\psi '}(pp)$ as a function of $x_L$ for
$\psi$'--productons at relevant RHIC energies, $\sqrt s=50$ and $500$GeV,
by using these spin--dependent gluon distributions together with
the spin--independent parton distribution of the Owens parametrization
\cite{Owens91} for (a), the BBS parametrizaton \cite{BBS95} for (b), and 
the GRV92 LO parametrization \cite{GRV92} for (c),
and taking $Q^2$ as $4m_c^2$ with $m_c=1.5$GeV.
Putting the ratio $\tilde\Theta/\Theta=1$,
we show the results of $A_{LL}^{\psi '}(pp)$ at $\sqrt s=50$
and $500$GeV in figs.2 and 3, respectively.
For $\sqrt s=50$GeV, we find that $A_{LL}^{\psi '}(pp)$ is very
effective for examining not only the magnitude
but also the behavior of the gluon polarization.
But at $\sqrt s=500$GeV we need rather precise data on
$A_{LL}^{\psi '}(pp)$ to test $\Delta g(x)$.
There is a tendency that the smaller the $\sqrt s$
is, the larger the $A_{LL}^{\psi '}(pp)$ become.
This is due to the fact that at larger $\sqrt s$, $x_a$ and $x_b$ ($=x_a-x_L$)
defined by eq.(\ref{eqn:def-x}) take smaller value, and so 
$g(x_a)$ and $g(x_b)$ at $\sqrt s=500$GeV become large.
Accodringly, $\Delta g(x_a)/g(x_a)\times\Delta g(x_b)/g(x_b)$ at
$\sqrt s=500$GeV become smaller than that at $\sqrt s=50$GeV.

In summary, observation of $A_{LL}^{\psi '}(pp)$ at small--$p_T$ regions
can largely contribute for the confirmation of the color--octet mechanism.
At present, we do not know the exact value of $\tilde\Theta/\Theta$ which
ranges in $0.69\div 1.63$.
It is interesting to fix this value from other experiments in order to
make a precise prediction of $A_{LL}^{\psi '}(pp)$.
Furthermore, the small--$p_T$ $\psi$'--production
allow us to give a rather clean test as a probe of the magnitude and
$x$--dependence of the gluon polarization.

\begin{center}
{\large \bf Figure captions}
\end{center}
\begin{description}
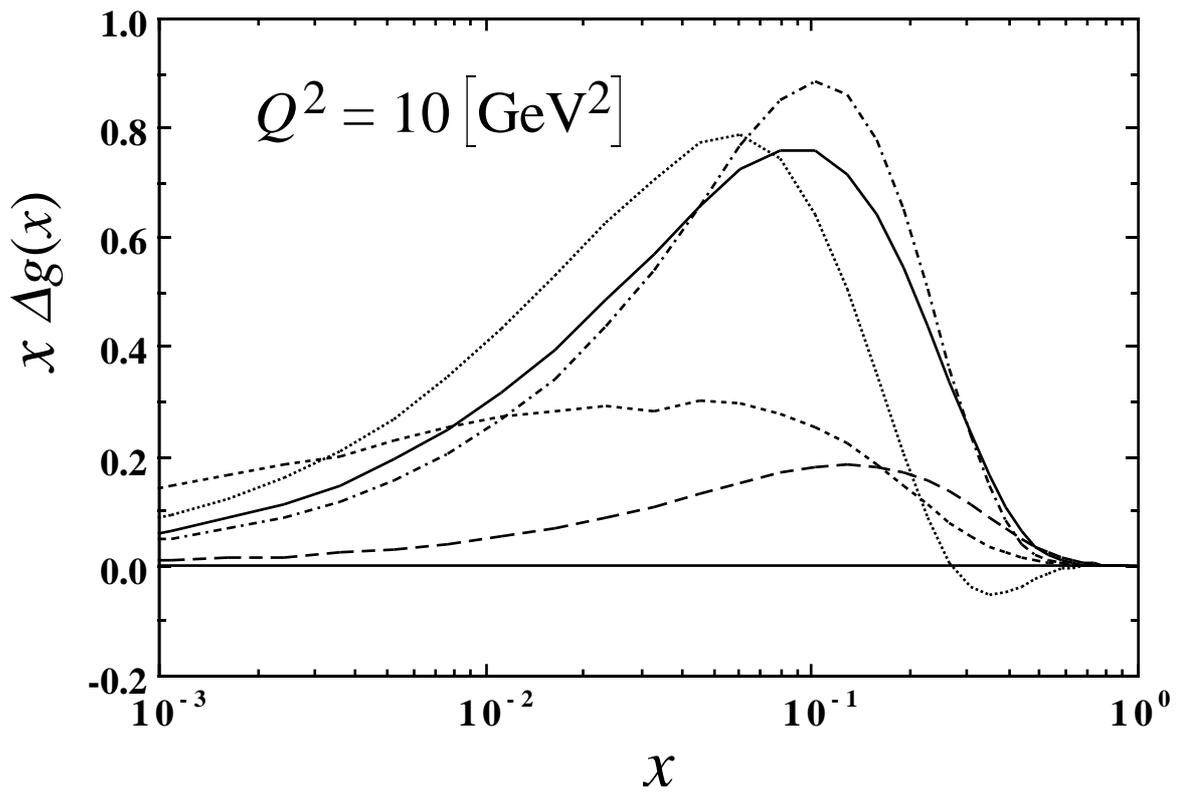
\item[Fig.\ 1:] The $x$ dependence of $x\Delta g(x, Q^2)$
at $Q^2=10$GeV$^2$ for various types of spin--dependent gluon distributions.
The solid, dash--dotted, dotted,
dashed and broken lines indicate the set A, B, C
of ref.\cite{GS95}, ref.\cite{BBS95} and the `standard senrio' of
ref.\cite{GRV95}, respectively.

\vspace{2em}

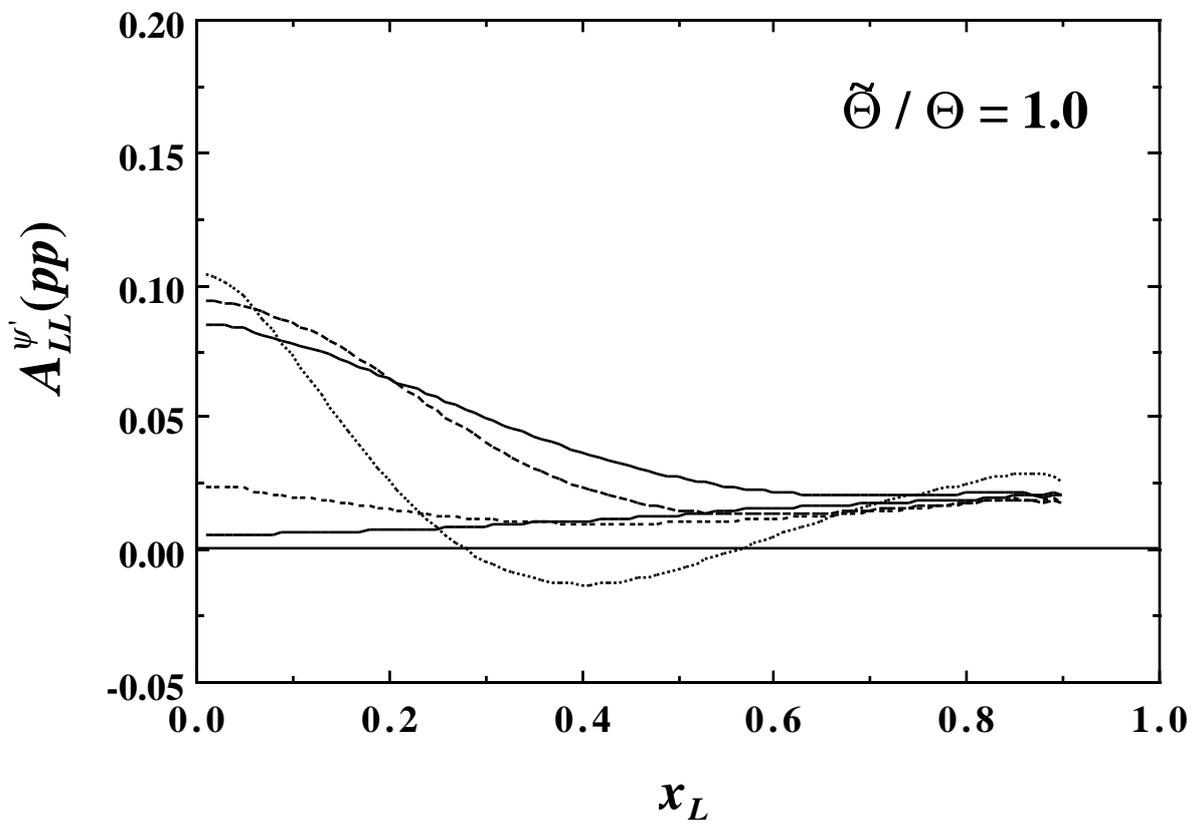
\item[Fig.\ 2:] The two--spin asymmetry $A_{LL}^{\psi '}(pp)$
for $\tilde\Theta/\Theta=1$ and $\sqrt s=50$GeV, calculated with various
types of $\Delta g(x)$, as a function of longitudinal momentum fraction $x_L$
of $\psi$'.
Various lines represent the same as in fig.1.

\vspace{2em}

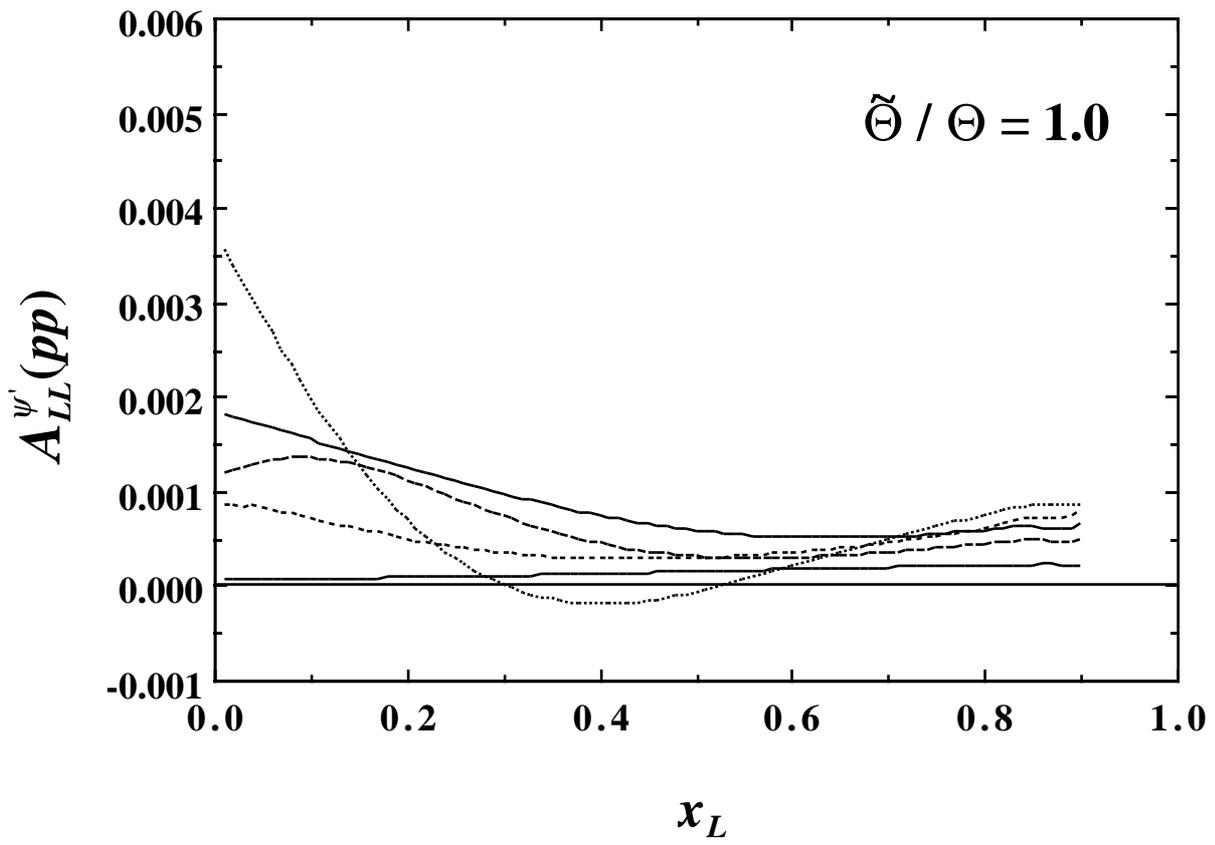
\item[Fig.\ 3:] The two--spin asymmetry $A_{LL}^{\psi '}(pp)$
for $\tilde\Theta/\Theta=1$ and $\sqrt s=500$GeV.
Various lines represent the same as in fig.1.
\end{description}


\begin{thebibliography}{99}
\bibitem{CDF}
F.\ Abe et.\ al., CDF Collab., Phys.\ Rev.\ Lett.\ {\bf 69} (1992) 3704;
Phys.\ Rev.\ Lett.\ {\bf 71} (1993) 2537.
\bibitem{octets}
E.\ Braaten and T.\ C.\ Yuan, Phys.\ Rev.\ Lett.\ {\bf 71} (1993) 1673;
P.\ Cho and A.\ K.\ Leibovich, Phys.\ Rev.\ {\bf D53} (1996) 150;
W.\ -K.\ Tang and M.\ V\"anttinen, Phys.\ Rev.\ {\bf D53} (1996) 4851;
S.\ Gupta and P.\ Mathews, preprint TIFR/TH/96--53, hep--ph/9609504 (1996).
\bibitem{Cacciari96}
M.\ Cacciari and M.\ Kr\"amer, Phys.\ Rev.\ Lett.\ {\bf 76} (1996) 4128.
\bibitem{pDIS}
J.\ Ashman et al., EMC Collab., Phys.\ Lett.\ {\bf B206} (1988) 364;
Nucl.\ Phys.\ {\bf B328} (1989) 1;
B.\ Adeva et al., SMC Collab., Phys.\ Lett.\ {\bf B302} (1993) 533;
{\bf B320} (1994) 400;
P.\ L.\ Anthony et al., E142 Collab., Phys.\ Rev.\ Lett.\ {\bf 71} (1993) 959;
D.\ Adeva et al., SMC Collab., Phys.\ Lett.\ {\bf B329} (1994) 399;
K.\ Abe et al., E143 Collab., Phys.\ Rev.\ Lett.\ {\bf 74} (1995) 346.
\bibitem{Beneke96}
M. Beneke and I.\ Z.\ Rothstein, Phys.\ Rev.\ {\bf D54} (1996) 2005.
\bibitem{Leibovich96}
M.\ V\"anttinen, preprint CALT--68--2082, hep--ph/9610381 (1996).
\bibitem{GS95}
T.\ Gehrmann and W.\ J.\ Stirling, Z.\ Phys.\ {\bf C65} (1995) 461.
\bibitem{BBS95}
S.\ J.\ Brodsky, M.\ Burkardt and I.\ Schmidt, Nucl.\ Phys.\ {\bf B441},
(1995) 197.
\bibitem{GRV95}
M.\ Gl\"uck, E.\ Reya and W.\ Vogelsang, Phys.\ Lett.\ {\bf B359} (1995) 201.
\bibitem{Owens91}
J.\ F.\ Owens, Phys.\ Lett.\ {\bf B266} (1991) 126.
\bibitem{GRV92}
M.\ Gl\"uck, E.\ Reya and W.\ Vogelsang, Z.\ Phys.\ {\bf C53} (1992) 127.
\end{thebibliography}
\end{document}